\documentclass{rrparticle}
\usepackage{graphicx}

\title{A quantum mechanical model for the rate of return} 
\author{Liviu-Adrian Cotfas}
\affil{Faculty of Economic Cybernetics, Statistics and Informatics, Academy of Economic Studies,\\ 6 Piata Romana, 010374 Bucharest, Romania, 
Email: {\em lcotfas@gmail.com}
}
\keywords{econophysics, quantum finance, finite quantum systems}
\pacs{89.65.Gh}

\begin{document}
\maketitle

\begin{abstract}
In their activity, the traders approximate the rate of return by integer multiples of a minimal one.
Therefore, it can be regarded as a quantized variable. On the other hand, there is the impossibility of observing the rate of return and its instantaneous forward time derivative, even if we consider it as a continuous variable. We present a quantum model for the rate of return based on the mathematical formalism used in the case of quantum systems with finite-dimensional Hilbert space. The rate of return is described by a discrete wave function and its time evolution by a Schr\" odinger type equation.
\end{abstract}

\section{Introduction}

The mathematical modeling of price dynamics in a stock market is a very complex problem [1-9]. A given stock has not a definite price until it is traded. The price is exactly known only at the time of sale when the stock is between the traders. We can never simultaneously know both the price of a stock and its owner [7-11].  On the other hand, there is the impossibility of observing prices and their instantaneous forward time derivative \cite{Segal}. The stock price $\wp $ is a discrete variable, not a continuous one. It is  an integer multiple of a certain minimal quantity, a sort of `quantum' of cash.
The rate of return  in a trading day 
\begin{equation}\label{ratedef}
\mathcal{R}=\frac{\wp -\wp _0}{\wp _0}
\end{equation}
where $\wp _0$ is the previous day's closing price, is also `quantized'.  The rate of return being directly related to the price, there is the impossibility of observing the rate of return  and its instantaneous forward time derivative.

The methods of quantum mechanics are among the tools used in order to get a deeper insight in the complexity of the financial markets. Our aim is to present a quantum model for the rate of  return. We shall take into consideration only the case of  stock markets with price limit rule: the rate of return in a trading day can not exceed a fixed limit $\pm q\%$ (for example, in most Chinese stock markets $q=10$). By choosing $1\%$ as a `quantum'  for the rate of return, the set of all the possible values of $\mathcal{R}$ is
\[
\mathcal{L}=\{ -q,\, -q\!+\!1,\, ...\, ,\, q\!-\!1,\, q\}.
\]

\section{A finite-dimensional quantum model }

Let $d\!=\!2q\!+\!1$. The space $\mathcal{H}$ of all the functions
\[
\psi  :\mathcal{L}\longrightarrow \mathbb{C}
\]
considered together with the scalar product 
\begin{equation}
\langle \psi ,\varphi \rangle =\sum_{n=-q}^q\overline{\psi (n)}\, \, \varphi (n)
\end{equation}
is a $d$-dimensional Hilbert space isomorphic to $\mathbb{C}^d$. For each function $\psi $ we use Dirac's notation $|\psi \rangle $ as an alternative notation, and the notation $\langle \psi |$ for the linear form
\[
\langle \psi | :\mathcal{H}\longrightarrow \mathbb{C}:|\varphi \rangle \mapsto \langle \psi |\varphi \rangle .
\]

The normalized function corresponding to $\psi \neq 0$, namely, 
\begin{equation}
\Psi  :\mathcal{L}\longrightarrow \mathbb{C}, \qquad \Psi (n)=\frac{1}{\sqrt{\langle \psi ,\psi \rangle }}\, \psi (n)
\end{equation}
satisfies the relation 
\begin{equation}
\sum_{n=-q}^q|\Psi (n)|^2=1.
\end{equation}
Following the analogy with the case of quantum systems \cite{Messiah,Vourdas}, we describe the `state of our financial system' by a normalized  function $\Psi  \!:\!\mathcal{L}\!\longrightarrow \!\mathbb{C}$
and consider that
\[
|\Psi (n)|^2
\]
represents the probability to have a return rate equal to $n\%$.

The set of functions $\{ |\delta _n\rangle \}_{n=-q}^q$, where
\begin{equation}
\delta _n  :\mathcal{L}\longrightarrow \mathbb{C},\qquad \delta _n(k)=\left\{ 
\begin{array}{lll}
1 & {\rm for} & k=n\\
0 & {\rm for} & k\neq n
\end{array}\right.
\end{equation}
is an orthonormal basis in $\mathcal{H}$, that is, we have the relations
\begin{equation}
\sum_{n=-q}^q|\delta _n\rangle \langle \delta _n|=\mathbb{I}\qquad {\rm and}\qquad \langle \delta _n|\delta _m\rangle =\left\{ 
\begin{array}{lll}
1 & {\rm for} & n=m\\
0 & {\rm for} & n\neq m
\end{array}\right.
\end{equation}
where $\mathbb{I} \!:\!\mathcal{H}\!\rightarrow \!\mathcal{H}$ is the identity operator $\mathbb{I}|\psi \rangle \!=\!|\psi \rangle $. Since $\langle \delta _n|\psi \rangle =\psi (n)$, we have
\begin{equation}
|\psi \rangle =\sum_{n=-q}^q\psi (n)\, |\delta _n\rangle \qquad {\rm and }\qquad 
\langle \psi | =\sum_{n=-q}^q\overline{\psi (n)}\, \langle \delta _n|.
\end{equation}
By following the analogy with the case of quantum systems \cite{Messiah,Vourdas}, we define the Hermitian operator corresponding to the {\em rate of return} as
\begin{equation}
\hat {\mathcal{R}}\!:\!\mathcal{H}\!\longrightarrow \!\mathcal{H},\qquad 
\hat {\mathcal{R}}=\sum_{n=-q}^{q}n\, |\delta _n\rangle \langle \delta _n|
\end{equation}
that is, in terms of functions, as
\begin{equation}
 \mathcal{H}\!\longrightarrow \!\mathcal{H}:\, \Psi \mapsto \hat {\mathcal{R}}\Psi  \qquad {\rm with}\qquad (\hat {\mathcal{R}}\Psi )(n)=n\, \Psi (n).
\end{equation}
The function $|\delta _n\rangle $ is an eigenfunction of $\hat {\mathcal{R}}$ corresponding to the eigenvalue $n$
\begin{equation}
\hat {\mathcal{R}}|\delta _n\rangle =n\, |\delta _n\rangle .
\end{equation}

The finite Fourier transform 
\begin{equation}
\mathcal{F}\!:\!\mathcal{H}\!\longrightarrow \!\mathcal{H},\qquad \mathcal{F}=\frac{1}{\sqrt{d}}\sum_{k,n=-q}^{q}{\rm e}^{-\frac{2\pi {\rm i}}{d}kn}\, |\delta _k\rangle \langle \delta _n|
\end{equation}
is a unitary transformation. Its inverse is the adjoint transformation
\begin{equation}
\mathcal{F}^+\!:\!\mathcal{H}\!\longrightarrow \!\mathcal{H},\qquad \mathcal{F}^+=\frac{1}{\sqrt{d}}\sum_{k,n=-q}^{q}{\rm e}^{\frac{2\pi {\rm i}}{d}kn}\, |\delta _k\rangle \langle \delta _n|
\end{equation}
and
\begin{equation}
|\psi \rangle =\mathcal{F}^+|\varphi \rangle \qquad \Longrightarrow \qquad \langle \psi |=\langle \varphi |\mathcal{F}.\\[3mm]
\end{equation}

The set of functions $\{ |\tilde \delta _n\rangle \}_{n=-q}^q$, where $|\tilde \delta _n\rangle =\mathcal{F}^+|\delta _n\rangle $, that is,
\begin{equation}
\begin{array}{l}
\tilde \delta _n  :\mathcal{L}\longrightarrow \mathbb{C},\qquad \tilde\delta _n(k)=\frac{1}{\sqrt{d}}\, {\rm e}^{\frac{2\pi {\rm i}}{d}kn}
\end{array}
\end{equation}
is an orthonormal basis in $\mathcal{H}$, that is, we have the relations
\begin{equation}
\sum_{n=-q}^q|\tilde \delta _n\rangle \langle \tilde\delta _n|=\mathbb{I}\qquad {\rm and}\qquad \langle \tilde \delta _n|\tilde \delta _m\rangle =\left\{ 
\begin{array}{lll}
1 & {\rm for} & n=m\\
0 & {\rm for} & n\neq m .
\end{array}\right.
\end{equation}
We define the operator corresponding to the {\em trend of the rate of return}, namely, 
\begin{equation}
\hat {\mathcal{T}}\!:\!\mathcal{H}\!\longrightarrow \!\mathcal{H},\qquad 
\hat {\mathcal{T}}=\sum_{n=-q}^{q}n\, |\tilde\delta _n\rangle \langle \tilde\delta _n|
\end{equation}
by following the analogy with the operator corresponding to the momentum in the case of the quantum systems with finite-dimensional Hilbert space \cite{Vourdas}. The function $|\tilde\delta _n\rangle $ is an eigenfunction of $\hat {\mathcal{T}}$ corresponding to the eigenvalue $n$
\begin{equation}
\hat {\mathcal{T}}|\tilde\delta _n\rangle =n\, |\tilde\delta _n\rangle 
\end{equation}
and
\[
\hat {\mathcal{T}}=\sum_{n=-q}^{q}n\, |\tilde\delta _n\rangle \langle \tilde\delta _n|=\sum_{n=-q}^{q}n\, \mathcal{F}^+|\delta _n\rangle \langle \delta _n|\mathcal{F}=\mathcal{F}^+\hat{\mathcal{R}}\mathcal{F}
\]
that is, 
\begin{equation}
\hat {\mathcal{T}}=\mathcal{F}^+\hat{\mathcal{R}}\mathcal{F}.
\end{equation}
In the case of a normalized function $\Psi $, the numbers
\[
\begin{array}{l}
\langle \hat {\mathcal{R}}\rangle =\langle \Psi , \hat {\mathcal{R}}\Psi \rangle =\sum\limits_{n=-q}^q\overline{\Psi (n)}\, \, \hat {\mathcal{R}}\Psi (n)=\sum\limits_{n=-q}^qn\, |\Psi (n)|^2\\
\langle \hat {\mathcal{T}}\rangle =\langle \Psi , \hat {\mathcal{T}}\Psi \rangle=\langle \Psi , \mathcal{F}^+\hat{\mathcal{R}}\mathcal{F}\Psi \rangle=\langle \mathcal{F}[ \Psi ], \hat{\mathcal{R}}\mathcal{F}[\Psi ]\rangle =\sum\limits_{n=-q}^qn\, | \mathcal{F}[ \Psi ](n)|^2
\end{array}
\]
represent the {\em mean value} of the rate of return and of the trend, respectively.

The other financial variables are also described by Hermitian operators. For example, the relation (\ref{ratedef}), written in the form
\[
\wp =\wp _0+\wp _0\, \mathcal{R}
\]
allows us to define the {\em price operator} as
\begin{equation}
\hat \wp =\wp _0\, \mathbb{I}+\wp _0\, \hat{\mathcal{R}}.
\end{equation}

\section{Time evolution of the rate of return}

By following the analogy with the quantum systems \cite{Messiah}, we assume that there exists a Hermitian operator of the form 
\begin{equation}\label{Hamiltonian}
\begin{array}{l}
\hat H=\frac{1}{2\mu }\hat{\mathcal{T}}^2+\mathcal{V}(\hat{\mathcal{R}},t)
\end{array}
\end{equation}
called the Hamiltonian ($\mu $ is a positive constant), such that the function 
\[
\mathcal{L}\!\times \!\mathbb{R}\!\longrightarrow \! \mathbb{C}:(n,t)\!\mapsto \!\Psi (n,t)
\]
 describing the time evolution of the state of our financial system satisfies the Schr\" odinger equation
\begin{equation}
{\rm i}\, \frac{\partial \Psi }{\partial t}=\hat H\, \Psi .
\end{equation}
The time dependent function $\Psi (n,t)$ is well-determined by its values $\Psi (n,0)$ at $t=0$.

The rate of return of the stock market in equilibrium is usually described by a Gaussian function \cite{Zhang}. The Jacobi theta function \cite{Ruzzi}
\begin{equation}
\theta _3(z,\tau )=\sum_{\alpha =-\infty }^\infty {\rm e}^{{\rm i}\pi \tau \alpha ^2}\, {\rm e}^{2\pi {\rm i}\alpha z}
\end{equation}
has several remarkable properties among which we mention:
\begin{equation}
\theta _3(z+1 ,\tau )= \theta _3(z,\tau )
\end{equation}
\begin{equation}
\theta _3(z,{\rm i}\tau )=\frac{1}{\sqrt{\tau }}\, {\rm exp}^{-\frac{\pi z^2}{\tau }}\, \theta _3 \left( \frac{z}{{\rm i}\tau },\frac{\rm i}{\tau }\right)
\end{equation}
and \cite{Ruzzi,Mehta}
\begin{equation}\label{Ruzzi}
\theta _3 \left( \frac{k}{d},\frac{{\rm i}\kappa }{d}\right)=\frac{1}{\sqrt{\kappa d}}\, \sum_{n=-q}^{q}{\rm e}^{-\frac{2\pi {\rm i}}{d}kn}\, \theta _3 \left( \frac{n}{d},\frac{\rm i}{ \kappa d}\right).
\end{equation}

For any $\kappa \in (0,\infty )$ the function 
\begin{equation}
\gamma _\kappa  :\mathcal{L}\longrightarrow \mathbb{C}, \qquad \gamma _\kappa  (n)=\sum _{m=-\infty }^\infty {\rm e}^{-\frac{\kappa \pi }{d}(md+n)^2}
\end{equation}
can be expressed in terms of the Jacobi function $\theta _3$ as
\begin{equation}
\gamma _\kappa (n)=\frac{1}{\sqrt{\kappa d}}\, \theta _3\left( \frac{n}{d},\frac{{\rm i}}{\kappa d} \right)
\end{equation}
and the relation (\ref{Ruzzi}) becomes
\begin{equation}
\mathcal{F}[\gamma _\kappa ]=\frac{1}{\sqrt{\kappa }}\, \gamma _{\frac{1}{\kappa }}.
\end{equation}
Since $||\gamma _{\frac{1}{\kappa }}||=\sqrt{\kappa }\, ||\mathcal{F}[\gamma _\kappa ]||=\sqrt{\kappa }\, ||\gamma _\kappa ||$, the normalized function 
\[
\Upsilon _\kappa :\mathcal{L}\longrightarrow \mathbb{C}, \qquad \Upsilon _\kappa (n)=\frac{\gamma _\kappa (n)}{||\gamma _\kappa ||}
\]
satisfies the relation
\begin{equation}
\mathcal{F}[\Upsilon _\kappa ]=\Upsilon _{\frac{1}{\kappa }}.
\end{equation}
The function $\Upsilon _\kappa $ can be regarded as a discrete version of the Gaussian function
(see Fig. \ref{gaussians}) depending on a parameter $\kappa \!\in \!(0,\infty )$.

\begin{center}
\begin{figure}[b]
\setlength{\unitlength}{1.8mm}
\begin{picture}(70,20)(-3,0)
\put(0,2){\vector(1,0){20}}
\put(23,2){\vector(1,0){20}}
\put(46,2){\vector(1,0){20}}
\put(9.5,0){\vector(0,1){17}}
\put(32.5,0){\vector(0,1){17}}
\put(55.5,0){\vector(0,1){17}}

\put(    1.50000,   2.45684){\circle*{0.3}}
\put(    2.30000,   2.60584){\circle*{0.3}}
\put(    3.10000,   2.91252){\circle*{0.3}}
\put(    3.90000,   3.38704){\circle*{0.3}}
\put(    4.70000,   4.02870){\circle*{0.3}}
\put(    5.50000,   4.81235){\circle*{0.3}}
\put(    6.30000,   5.67880){\circle*{0.3}}
\put(    7.10000,   6.53499){\circle*{0.3}}
\put(    7.90000,   7.26650){\circle*{0.3}}
\put(    8.70000,   7.76099){\circle*{0.3}}
\put(    9.50000,   7.93595){\circle*{0.3}}
\put(   10.30000,   7.76099){\circle*{0.3}}
\put(   11.10000,   7.26650){\circle*{0.3}}
\put(   11.90000,   6.53499){\circle*{0.3}}
\put(   12.70000,   5.67880){\circle*{0.3}}
\put(   13.50000,   4.81235){\circle*{0.3}}
\put(   14.30000,   4.02870){\circle*{0.3}}
\put(   15.10000,   3.38704){\circle*{0.3}}
\put(   15.90000,   2.91252){\circle*{0.3}}
\put(   16.70000,   2.60584){\circle*{0.3}}
\put(   17.50000,   2.45684){\circle*{0.3}}
\put(    1.50000,2){\line(0,1){    .45684}}
\put(    2.30000,2){\line(0,1){    .60584}}
\put(    3.10000,2){\line(0,1){    .91252}}
\put(    3.90000,2){\line(0,1){   1.38704}}
\put(    4.70000,2){\line(0,1){   2.02870}}
\put(    5.50000,2){\line(0,1){   2.81235}}
\put(    6.30000,2){\line(0,1){   3.67880}}
\put(    7.10000,2){\line(0,1){   4.53499}}
\put(    7.90000,2){\line(0,1){   5.26650}}
\put(    8.70000,2){\line(0,1){   5.76099}}
\put(    9.50000,2){\line(0,1){   5.93595}}
\put(   10.30000,2){\line(0,1){   5.76099}}
\put(   11.10000,2){\line(0,1){   5.26650}}
\put(   11.90000,2){\line(0,1){   4.53499}}
\put(   12.70000,2){\line(0,1){   3.67880}}
\put(   13.50000,2){\line(0,1){   2.81235}}
\put(   14.30000,2){\line(0,1){   2.02870}}
\put(   15.10000,2){\line(0,1){   1.38704}}
\put(   15.90000,2){\line(0,1){    .91252}}
\put(   16.70000,2){\line(0,1){    .60584}}
\put(   17.50000,2){\line(0,1){    .45684}}
\put(   24.50000,   2.00000){\circle*{0.3}}
\put(   25.30000,   2.00000){\circle*{0.3}}
\put(   26.10000,   2.00062){\circle*{0.3}}
\put(   26.90000,   2.00582){\circle*{0.3}}
\put(   27.70000,   2.04073){\circle*{0.3}}
\put(   28.50000,   2.21114){\circle*{0.3}}
\put(   29.30000,   2.81152){\circle*{0.3}}
\put(   30.10000,   4.31254){\circle*{0.3}}
\put(   30.90000,   6.88587){\circle*{0.3}}
\put(   31.70000,   9.65336){\circle*{0.3}}
\put(   32.50000,  10.88838){\circle*{0.3}}
\put(   33.30000,   9.65336){\circle*{0.3}}
\put(   34.10000,   6.88587){\circle*{0.3}}
\put(   34.90000,   4.31254){\circle*{0.3}}
\put(   35.70000,   2.81152){\circle*{0.3}}
\put(   36.50000,   2.21114){\circle*{0.3}}
\put(   37.30000,   2.04073){\circle*{0.3}}
\put(   38.10000,   2.00582){\circle*{0.3}}
\put(   38.90000,   2.00062){\circle*{0.3}}
\put(   39.70000,   2.00000){\circle*{0.3}}
\put(   40.50000,   2.00000){\circle*{0.3}}
\put(   24.50000,2){\line(0,1){    .00000}}
\put(   25.30000,2){\line(0,1){    .00000}}
\put(   26.10000,2){\line(0,1){    .00062}}
\put(   26.90000,2){\line(0,1){    .00582}}
\put(   27.70000,2){\line(0,1){    .04073}}
\put(   28.50000,2){\line(0,1){    .21114}}
\put(   29.30000,2){\line(0,1){    .81152}}
\put(   30.10000,2){\line(0,1){   2.31254}}
\put(   30.90000,2){\line(0,1){   4.88587}}
\put(   31.70000,2){\line(0,1){   7.65336}}
\put(   32.50000,2){\line(0,1){   8.88838}}
\put(   33.30000,2){\line(0,1){   7.65336}}
\put(   34.10000,2){\line(0,1){   4.88587}}
\put(   34.90000,2){\line(0,1){   2.31254}}
\put(   35.70000,2){\line(0,1){    .81152}}
\put(   36.50000,2){\line(0,1){    .21114}}
\put(   37.30000,2){\line(0,1){    .04073}}
\put(   38.10000,2){\line(0,1){    .00582}}
\put(   38.90000,2){\line(0,1){    .00062}}
\put(   39.70000,2){\line(0,1){    .00000}}
\put(   40.50000,2){\line(0,1){    .00000}}
\put(   47.50000,   2.00000){\circle*{0.3}}
\put(   48.30000,   2.00000){\circle*{0.3}}
\put(   49.10000,   2.00000){\circle*{0.3}}
\put(   49.90000,   2.00000){\circle*{0.3}}
\put(   50.70000,   2.00022){\circle*{0.3}}
\put(   51.50000,   2.00596){\circle*{0.3}}
\put(   52.30000,   2.08811){\circle*{0.3}}
\put(   53.10000,   2.71551){\circle*{0.3}}
\put(   53.90000,   5.19387){\circle*{0.3}}
\put(   54.70000,   9.83682){\circle*{0.3}}
\put(   55.50000,  12.57013){\circle*{0.3}}
\put(   56.30000,   9.83682){\circle*{0.3}}
\put(   57.10000,   5.19387){\circle*{0.3}}
\put(   57.90000,   2.71551){\circle*{0.3}}
\put(   58.70000,   2.08811){\circle*{0.3}}
\put(   59.50000,   2.00596){\circle*{0.3}}
\put(   60.30000,   2.00022){\circle*{0.3}}
\put(   61.10000,   2.00000){\circle*{0.3}}
\put(   61.90000,   2.00000){\circle*{0.3}}
\put(   62.70000,   2.00000){\circle*{0.3}}
\put(   63.50000,   2.00000){\circle*{0.3}}
\put(   47.50000,2){\line(0,1){    .00000}}
\put(   48.30000,2){\line(0,1){    .00000}}
\put(   49.10000,2){\line(0,1){    .00000}}
\put(   49.90000,2){\line(0,1){    .00000}}
\put(   50.70000,2){\line(0,1){    .00022}}
\put(   51.50000,2){\line(0,1){    .00596}}
\put(   52.30000,2){\line(0,1){    .08811}}
\put(   53.10000,2){\line(0,1){    .71551}}
\put(   53.90000,2){\line(0,1){   3.19387}}
\put(   54.70000,2){\line(0,1){   7.83682}}
\put(   55.50000,2){\line(0,1){  10.57013}}
\put(   56.30000,2){\line(0,1){   7.83682}}
\put(   57.10000,2){\line(0,1){   3.19387}}
\put(   57.90000,2){\line(0,1){    .71551}}
\put(   58.70000,2){\line(0,1){    .08811}}
\put(   59.50000,2){\line(0,1){    .00596}}
\put(   60.30000,2){\line(0,1){    .00022}}
\put(   61.10000,2){\line(0,1){    .00000}}
\put(   61.90000,2){\line(0,1){    .00000}}
\put(   62.70000,2){\line(0,1){    .00000}}
\put(   63.50000,2){\line(0,1){    .00000}}


\put(0,0.9){$\scriptscriptstyle{-10}$}
\put(17,0.9){$\scriptscriptstyle{10}$}
\put(23,0.9){$\scriptscriptstyle{-10}$}
\put(40,0.9){$\scriptscriptstyle{10}$}
\put(46,0.9){$\scriptscriptstyle{-10}$}
\put(63,0.9){$\scriptscriptstyle{10}$}
\put(4.5,0.9){$\scriptscriptstyle{-5}$}
\put(13.3,0.9){$\scriptscriptstyle{5}$}
\put(27.5,0.9){$\scriptscriptstyle{-5}$}
\put(36.3,0.9){$\scriptscriptstyle{5}$}
\put(50.5,0.9){$\scriptscriptstyle{-5}$}
\put(59.3,0.9){$\scriptscriptstyle{5}$}

\put(6.7,13.5){$\scriptscriptstyle{0.75}$}
\put( 9.3,14){\line(1,0){0.4}}
\put(7.2,9.5){$\scriptscriptstyle{0.5}$}
\put( 9.3,10){\line(1,0){0.4}}
\put(19,2.5){$\scriptstyle{n}$}
\put(10,16){$\scriptstyle{\gamma _{0.2}(n)}$}

\put(29.7,13.5){$\scriptscriptstyle{0.75}$}
\put( 32.3,14){\line(1,0){0.4}}
\put(42,2.5){$\scriptstyle{n}$}
\put(33,16){$\scriptstyle{\gamma _{1}(n)}$}

\put(52.7,13.5){$\scriptscriptstyle{0.75}$}
\put( 55.3,14){\line(1,0){0.4}}
\put(65,2.5){$\scriptstyle{n}$}
\put(56,16){$\scriptstyle{\gamma _{2}(n)}$}
\end{picture}
\caption{The functions $\gamma _{0.2}$ (left),  $\gamma _1$ (center) and $\gamma _2$ (right) in the case $q=10$.}\label{gaussians}
\end{figure}
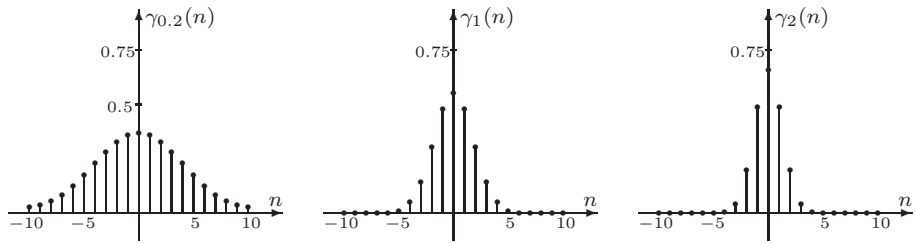
\end{center}
\begin{center}
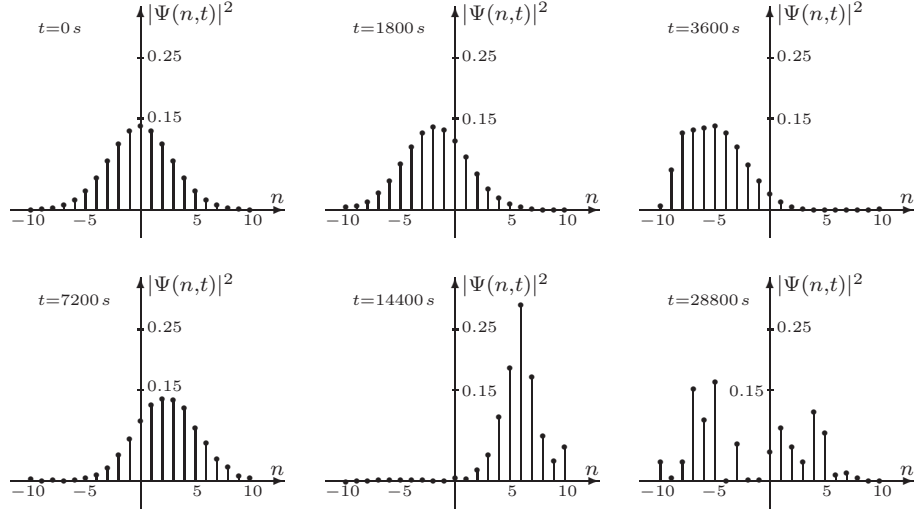
\begin{figure}[t]
\setlength{\unitlength}{1.8mm}
\begin{picture}(70,40)(-3,0)
\put(0,2){\vector(1,0){20}}
\put(23,2){\vector(1,0){20}}
\put(46,2){\vector(1,0){20}}
\put(9.5,0){\vector(0,1){17}}
\put(32.5,0){\vector(0,1){17}}
\put(55.5,0){\vector(0,1){17}}

\put(0,22){\vector(1,0){20}}
\put(23,22){\vector(1,0){20}}
\put(46,22){\vector(1,0){20}}
\put(9.5,20){\vector(0,1){17}}
\put(32.5,20){\vector(0,1){17}}
\put(55.5,20){\vector(0,1){17}}

\put(    1.50000,  22.03669){\circle*{0.3}}
\put(    2.30000,  22.06452){\circle*{0.3}}
\put(    3.10000,  22.14637){\circle*{0.3}}
\put(    3.90000,  22.33818){\circle*{0.3}}
\put(    4.70000,  22.72346){\circle*{0.3}}
\put(    5.50000,  23.39031){\circle*{0.3}}
\put(    6.30000,  24.37895){\circle*{0.3}}
\put(    7.10000,  25.61516){\circle*{0.3}}
\put(    7.90000,  26.87548){\circle*{0.3}}
\put(    8.70000,  27.83402){\circle*{0.3}}
\put(    9.50000,  28.19375){\circle*{0.3}}
\put(   10.30000,  27.83402){\circle*{0.3}}
\put(   11.10000,  26.87548){\circle*{0.3}}
\put(   11.90000,  25.61516){\circle*{0.3}}
\put(   12.70000,  24.37895){\circle*{0.3}}
\put(   13.50000,  23.39031){\circle*{0.3}}
\put(   14.30000,  22.72346){\circle*{0.3}}
\put(   15.10000,  22.33818){\circle*{0.3}}
\put(   15.90000,  22.14637){\circle*{0.3}}
\put(   16.70000,  22.06452){\circle*{0.3}}
\put(   17.50000,  22.03669){\circle*{0.3}}
\put(    1.50000,22){\line(0,1){    .03669}}
\put(    2.30000,22){\line(0,1){    .06452}}
\put(    3.10000,22){\line(0,1){    .14637}}
\put(    3.90000,22){\line(0,1){    .33818}}
\put(    4.70000,22){\line(0,1){    .72346}}
\put(    5.50000,22){\line(0,1){   1.39031}}
\put(    6.30000,22){\line(0,1){   2.37895}}
\put(    7.10000,22){\line(0,1){   3.61516}}
\put(    7.90000,22){\line(0,1){   4.87548}}
\put(    8.70000,22){\line(0,1){   5.83403}}
\put(    9.50000,22){\line(0,1){   6.19376}}
\put(   10.30000,22){\line(0,1){   5.83403}}
\put(   11.10000,22){\line(0,1){   4.87548}}
\put(   11.90000,22){\line(0,1){   3.61516}}
\put(   12.70000,22){\line(0,1){   2.37895}}
\put(   13.50000,22){\line(0,1){   1.39031}}
\put(   14.30000,22){\line(0,1){    .72346}}
\put(   15.10000,22){\line(0,1){    .33818}}
\put(   15.90000,22){\line(0,1){    .14637}}
\put(   16.70000,22){\line(0,1){    .06452}}
\put(   17.50000,22){\line(0,1){    .03669}}
\put(   24.50000,  22.22564){\circle*{0.3}}
\put(   25.30000,  22.31724){\circle*{0.3}}
\put(   26.10000,  22.58692){\circle*{0.3}}
\put(   26.90000,  23.27954){\circle*{0.3}}
\put(   27.70000,  24.14642){\circle*{0.3}}
\put(   28.50000,  25.35655){\circle*{0.3}}
\put(   29.30000,  26.60616){\circle*{0.3}}
\put(   30.10000,  27.64395){\circle*{0.3}}
\put(   30.90000,  28.13188){\circle*{0.3}}
\put(   31.70000,  27.92664){\circle*{0.3}}
\put(   32.50000,  27.08824){\circle*{0.3}}
\put(   33.30000,  25.88869){\circle*{0.3}}
\put(   34.10000,  24.63945){\circle*{0.3}}
\put(   34.90000,  23.58545){\circle*{0.3}}
\put(   35.70000,  22.84578){\circle*{0.3}}
\put(   36.50000,  22.41966){\circle*{0.3}}
\put(   37.30000,  22.20502){\circle*{0.3}}
\put(   38.10000,  22.08581){\circle*{0.3}}
\put(   38.90000,  22.01937){\circle*{0.3}}
\put(   39.70000,  22.00134){\circle*{0.3}}
\put(   40.50000,  22.00000){\circle*{0.3}}
\put(   24.50000,22){\line(0,1){    .22564}}
\put(   25.30000,22){\line(0,1){    .31724}}
\put(   26.10000,22){\line(0,1){    .58692}}
\put(   26.90000,22){\line(0,1){   1.27954}}
\put(   27.70000,22){\line(0,1){   2.14642}}
\put(   28.50000,22){\line(0,1){   3.35655}}
\put(   29.30000,22){\line(0,1){   4.60615}}
\put(   30.10000,22){\line(0,1){   5.64395}}
\put(   30.90000,22){\line(0,1){   6.13188}}
\put(   31.70000,22){\line(0,1){   5.92664}}
\put(   32.50000,22){\line(0,1){   5.08824}}
\put(   33.30000,22){\line(0,1){   3.88869}}
\put(   34.10000,22){\line(0,1){   2.63945}}
\put(   34.90000,22){\line(0,1){   1.58545}}
\put(   35.70000,22){\line(0,1){    .84578}}
\put(   36.50000,22){\line(0,1){    .41966}}
\put(   37.30000,22){\line(0,1){    .20503}}
\put(   38.10000,22){\line(0,1){    .08581}}
\put(   38.90000,22){\line(0,1){    .01937}}
\put(   39.70000,22){\line(0,1){    .00134}}
\put(   40.50000,22){\line(0,1){    .00000}}
\put(   47.50000,  22.26967){\circle*{0.3}}
\put(   48.30000,  24.96190){\circle*{0.3}}
\put(   49.10000,  27.66199){\circle*{0.3}}
\put(   49.90000,  27.85063){\circle*{0.3}}
\put(   50.70000,  28.02537){\circle*{0.3}}
\put(   51.50000,  28.21882){\circle*{0.3}}
\put(   52.30000,  27.69331){\circle*{0.3}}
\put(   53.10000,  26.65768){\circle*{0.3}}
\put(   53.90000,  25.35196){\circle*{0.3}}
\put(   54.70000,  24.12429){\circle*{0.3}}
\put(   55.50000,  23.17502){\circle*{0.3}}
\put(   56.30000,  22.56843){\circle*{0.3}}
\put(   57.10000,  22.24021){\circle*{0.3}}
\put(   57.90000,  22.09026){\circle*{0.3}}
\put(   58.70000,  22.03062){\circle*{0.3}}
\put(   59.50000,  22.00964){\circle*{0.3}}
\put(   60.30000,  22.00306){\circle*{0.3}}
\put(   61.10000,  22.00177){\circle*{0.3}}
\put(   61.90000,  22.00281){\circle*{0.3}}
\put(   62.70000,  22.01045){\circle*{0.3}}
\put(   63.50000,  22.05211){\circle*{0.3}}
\put(   47.50000,22){\line(0,1){    .26967}}
\put(   48.30000,22){\line(0,1){   2.96190}}
\put(   49.10000,22){\line(0,1){   5.66199}}
\put(   49.90000,22){\line(0,1){   5.85063}}
\put(   50.70000,22){\line(0,1){   6.02537}}
\put(   51.50000,22){\line(0,1){   6.21882}}
\put(   52.30000,22){\line(0,1){   5.69331}}
\put(   53.10000,22){\line(0,1){   4.65768}}
\put(   53.90000,22){\line(0,1){   3.35196}}
\put(   54.70000,22){\line(0,1){   2.12429}}
\put(   55.50000,22){\line(0,1){   1.17502}}
\put(   56.30000,22){\line(0,1){    .56843}}
\put(   57.10000,22){\line(0,1){    .24021}}
\put(   57.90000,22){\line(0,1){    .09026}}
\put(   58.70000,22){\line(0,1){    .03062}}
\put(   59.50000,22){\line(0,1){    .00964}}
\put(   60.30000,22){\line(0,1){    .00306}}
\put(   61.10000,22){\line(0,1){    .00177}}
\put(   61.90000,22){\line(0,1){    .00281}}
\put(   62.70000,22){\line(0,1){    .01045}}
\put(   63.50000,22){\line(0,1){    .05211}}
\put(    1.50000,   2.15477){\circle*{0.3}}
\put(    2.30000,   2.07101){\circle*{0.3}}
\put(    3.10000,   2.09125){\circle*{0.3}}
\put(    3.90000,   2.05141){\circle*{0.3}}
\put(    4.70000,   2.14277){\circle*{0.3}}
\put(    5.50000,   2.22563){\circle*{0.3}}
\put(    6.30000,   2.47562){\circle*{0.3}}
\put(    7.10000,   3.01863){\circle*{0.3}}
\put(    7.90000,   3.92814){\circle*{0.3}}
\put(    8.70000,   5.13281){\circle*{0.3}}
\put(    9.50000,   6.46502){\circle*{0.3}}
\put(   10.30000,   7.60246){\circle*{0.3}}
\put(   11.10000,   8.05389){\circle*{0.3}}
\put(   11.90000,   7.98761){\circle*{0.3}}
\put(   12.70000,   7.41350){\circle*{0.3}}
\put(   13.50000,   5.93488){\circle*{0.3}}
\put(   14.30000,   4.85403){\circle*{0.3}}
\put(   15.10000,   3.67411){\circle*{0.3}}
\put(   15.90000,   3.04326){\circle*{0.3}}
\put(   16.70000,   2.44318){\circle*{0.3}}
\put(   17.50000,   2.23601){\circle*{0.3}}
\put(    1.50000,2){\line(0,1){    .15477}}
\put(    2.30000,2){\line(0,1){    .07101}}
\put(    3.10000,2){\line(0,1){    .09125}}
\put(    3.90000,2){\line(0,1){    .05141}}
\put(    4.70000,2){\line(0,1){    .14277}}
\put(    5.50000,2){\line(0,1){    .22563}}
\put(    6.30000,2){\line(0,1){    .47562}}
\put(    7.10000,2){\line(0,1){   1.01863}}
\put(    7.90000,2){\line(0,1){   1.92814}}
\put(    8.70000,2){\line(0,1){   3.13281}}
\put(    9.50000,2){\line(0,1){   4.46502}}
\put(   10.30000,2){\line(0,1){   5.60246}}
\put(   11.10000,2){\line(0,1){   6.05390}}
\put(   11.90000,2){\line(0,1){   5.98761}}
\put(   12.70000,2){\line(0,1){   5.41350}}
\put(   13.50000,2){\line(0,1){   3.93488}}
\put(   14.30000,2){\line(0,1){   2.85403}}
\put(   15.10000,2){\line(0,1){   1.67411}}
\put(   15.90000,2){\line(0,1){   1.04326}}
\put(   16.70000,2){\line(0,1){    .44318}}
\put(   17.50000,2){\line(0,1){    .23601}}
\put(   24.50000,   2.00136){\circle*{0.3}}
\put(   25.30000,   2.01125){\circle*{0.3}}
\put(   26.10000,   2.02640){\circle*{0.3}}
\put(   26.90000,   2.08843){\circle*{0.3}}
\put(   27.70000,   2.11048){\circle*{0.3}}
\put(   28.50000,   2.11762){\circle*{0.3}}
\put(   29.30000,   2.10123){\circle*{0.3}}
\put(   30.10000,   2.09269){\circle*{0.3}}
\put(   30.90000,   2.04075){\circle*{0.3}}
\put(   31.70000,   2.05542){\circle*{0.3}}
\put(   32.50000,   2.23441){\circle*{0.3}}
\put(   33.30000,   2.19161){\circle*{0.3}}
\put(   34.10000,   2.82174){\circle*{0.3}}
\put(   34.90000,   3.91892){\circle*{0.3}}
\put(   35.70000,   6.71830){\circle*{0.3}}
\put(   36.50000,  10.36617){\circle*{0.3}}
\put(   37.30000,  15.01297){\circle*{0.3}}
\put(   38.10000,   9.71412){\circle*{0.3}}
\put(   38.90000,   5.34343){\circle*{0.3}}
\put(   39.70000,   3.49206){\circle*{0.3}}
\put(   40.50000,   4.54061){\circle*{0.3}}
\put(   24.50000,2){\line(0,1){    .00136}}
\put(   25.30000,2){\line(0,1){    .01125}}
\put(   26.10000,2){\line(0,1){    .02640}}
\put(   26.90000,2){\line(0,1){    .08843}}
\put(   27.70000,2){\line(0,1){    .11048}}
\put(   28.50000,2){\line(0,1){    .11762}}
\put(   29.30000,2){\line(0,1){    .10123}}
\put(   30.10000,2){\line(0,1){    .09269}}
\put(   30.90000,2){\line(0,1){    .04075}}
\put(   31.70000,2){\line(0,1){    .05542}}
\put(   32.50000,2){\line(0,1){    .23441}}
\put(   33.30000,2){\line(0,1){    .19161}}
\put(   34.10000,2){\line(0,1){    .82175}}
\put(   34.90000,2){\line(0,1){   1.91892}}
\put(   35.70000,2){\line(0,1){   4.71830}}
\put(   36.50000,2){\line(0,1){   8.36617}}
\put(   37.30000,2){\line(0,1){  13.01297}}
\put(   38.10000,2){\line(0,1){   7.71413}}
\put(   38.90000,2){\line(0,1){   3.34343}}
\put(   39.70000,2){\line(0,1){   1.49206}}
\put(   40.50000,2){\line(0,1){   2.54061}}
\put(   47.50000,   3.45071){\circle*{0.3}}
\put(   48.30000,   2.25480){\circle*{0.3}}
\put(   49.10000,   3.44982){\circle*{0.3}}
\put(   49.90000,   8.82281){\circle*{0.3}}
\put(   50.70000,   6.50720){\circle*{0.3}}
\put(   51.50000,   9.35129){\circle*{0.3}}
\put(   52.30000,   2.06901){\circle*{0.3}}
\put(   53.10000,   4.77297){\circle*{0.3}}
\put(   53.90000,   2.08961){\circle*{0.3}}
\put(   54.70000,   2.08911){\circle*{0.3}}
\put(   55.50000,   4.17956){\circle*{0.3}}
\put(   56.30000,   5.93645){\circle*{0.3}}
\put(   57.10000,   4.53066){\circle*{0.3}}
\put(   57.90000,   3.39991){\circle*{0.3}}
\put(   58.70000,   7.12186){\circle*{0.3}}
\put(   59.50000,   5.55734){\circle*{0.3}}
\put(   60.30000,   2.49856){\circle*{0.3}}
\put(   61.10000,   2.62317){\circle*{0.3}}
\put(   61.90000,   2.24171){\circle*{0.3}}
\put(   62.70000,   2.03045){\circle*{0.3}}
\put(   63.50000,   2.02298){\circle*{0.3}}
\put(   47.50000,2){\line(0,1){   1.45071}}
\put(   48.30000,2){\line(0,1){    .25480}}
\put(   49.10000,2){\line(0,1){   1.44982}}
\put(   49.90000,2){\line(0,1){   6.82281}}
\put(   50.70000,2){\line(0,1){   4.50720}}
\put(   51.50000,2){\line(0,1){   7.35129}}
\put(   52.30000,2){\line(0,1){    .06901}}
\put(   53.10000,2){\line(0,1){   2.77297}}
\put(   53.90000,2){\line(0,1){    .08961}}
\put(   54.70000,2){\line(0,1){    .08911}}
\put(   55.50000,2){\line(0,1){   2.17956}}
\put(   56.30000,2){\line(0,1){   3.93645}}
\put(   57.10000,2){\line(0,1){   2.53066}}
\put(   57.90000,2){\line(0,1){   1.39991}}
\put(   58.70000,2){\line(0,1){   5.12186}}
\put(   59.50000,2){\line(0,1){   3.55734}}
\put(   60.30000,2){\line(0,1){    .49856}}
\put(   61.10000,2){\line(0,1){    .62317}}
\put(   61.90000,2){\line(0,1){    .24171}}
\put(   62.70000,2){\line(0,1){    .03045}}
\put(   63.50000,2){\line(0,1){    .02298}}

\put(0,20.9){$\scriptscriptstyle{-10}$}
\put(17,20.9){$\scriptscriptstyle{10}$}
\put(23,20.9){$\scriptscriptstyle{-10}$}
\put(40,20.9){$\scriptscriptstyle{10}$}
\put(46,20.9){$\scriptscriptstyle{-10}$}
\put(63,20.9){$\scriptscriptstyle{10}$}

\put(2,35){$\scriptscriptstyle{t=0\, s}$}
\put(10,33){$\scriptscriptstyle{0.25}$}
\put( 9.3,33.2){\line(1,0){0.4}}
\put(10,28.5){$\scriptscriptstyle{0.15}$}
\put( 9.3,28.75){\line(1,0){0.4}}
\put(19,22.5){$\scriptstyle{n}$}
\put(10,36){$\scriptstyle{|\Psi(n,t)|^2}$}

\put(25,35){$\scriptscriptstyle{t=1800\, s}$}
\put(33,33){$\scriptscriptstyle{0.25}$}
\put( 32.3,33.2){\line(1,0){0.4}}
\put(33,28.3){$\scriptscriptstyle{0.15}$}
\put( 32.3,28.75){\line(1,0){0.4}}
\put(42,22.5){$\scriptstyle{n}$}
\put(33,36){$\scriptstyle{|\Psi(n,t)|^2}$}

\put(48,35){$\scriptscriptstyle{t=3600\, s}$}
\put(56,33){$\scriptscriptstyle{0.25}$}
\put( 55.3,33.2){\line(1,0){0.4}}
\put(56,28.3){$\scriptscriptstyle{0.15}$}
\put( 55.3,28.75){\line(1,0){0.4}}
\put(65,22.5){$\scriptstyle{n}$}
\put(56,36){$\scriptstyle{|\Psi(n,t)|^2}$}
\put(0,0.9){$\scriptscriptstyle{-10}$}
\put(17,0.9){$\scriptscriptstyle{10}$}
\put(23,0.9){$\scriptscriptstyle{-10}$}
\put(40,0.9){$\scriptscriptstyle{10}$}
\put(46,0.9){$\scriptscriptstyle{-10}$}
\put(63,0.9){$\scriptscriptstyle{10}$}
\put(4.5,0.9){$\scriptscriptstyle{-5}$}
\put(13.3,0.9){$\scriptscriptstyle{5}$}
\put(27.5,0.9){$\scriptscriptstyle{-5}$}
\put(36.3,0.9){$\scriptscriptstyle{5}$}
\put(50.5,0.9){$\scriptscriptstyle{-5}$}
\put(59.3,0.9){$\scriptscriptstyle{5}$}

\put(4.5,20.9){$\scriptscriptstyle{-5}$}
\put(13.3,20.9){$\scriptscriptstyle{5}$}
\put(27.5,20.9){$\scriptscriptstyle{-5}$}
\put(36.3,20.9){$\scriptscriptstyle{5}$}
\put(50.5,20.9){$\scriptscriptstyle{-5}$}
\put(59.3,20.9){$\scriptscriptstyle{5}$}

\put(2,15){$\scriptscriptstyle{t=7200\, s}$}
\put(10,13){$\scriptscriptstyle{0.25}$}
\put( 9.3,13.2){\line(1,0){0.4}}
\put(10,8.7){$\scriptscriptstyle{0.15}$}
\put( 9.3,8.75){\line(1,0){0.4}}
\put(19,2.5){$\scriptstyle{n}$}
\put(10,16){$\scriptstyle{|\Psi(n,t)|^2}$}

\put(25,15){$\scriptscriptstyle{t=14400\, s}$}
\put(33,13){$\scriptscriptstyle{0.25}$}
\put( 32.3,13.2){\line(1,0){0.4}}
\put(33,8.3){$\scriptscriptstyle{0.15}$}
\put( 32.3,8.75){\line(1,0){0.4}}
\put(42,2.5){$\scriptstyle{n}$}
\put(33,16){$\scriptstyle{|\Psi(n,t)|^2}$}

\put(48,15){$\scriptscriptstyle{t=28800\, s}$}
\put(56,13){$\scriptscriptstyle{0.25}$}
\put( 55.3,13.2){\line(1,0){0.4}}
\put(52.5,8.3){$\scriptscriptstyle{0.15}$}
\put( 55.3,8.75){\line(1,0){0.4}}
\put(65,2.5){$\scriptstyle{n}$}
\put(56,16){$\scriptstyle{|\Psi(n,t)|^2}$}
\end{picture}
\caption{The probability $|\Psi(n,t)|^2$ to have a rate of return equal to $n\%$ for $t\!=\!0$, 1800, 3600, 7200, 14400 and  28800 in the particular case $q\!=\!10$, $\kappa \!=\!0.2$, $\mu \!=\!1$, $\beta \!=\!1/10$, $\omega \!=\!1/5000$.}
\end{figure}
\end{center}

In the case of a Hamiltonian (\ref{Hamiltonian}), 
the kinetic part $\frac{1}{2\mu }\hat{\mathcal{T}}^2$ represents the efforts of the traders to change prices \cite{Choustova}. The intensive exchange of information in the world of finances is one of the main sources determining dynamics of prices. The potential part $\mathcal{V}(\hat {\mathcal{R}},t)$ of $\hat H$ describes the interactions between traders as well as external economic conditions and even meteorological conditions \cite{Choustova}.

The total effect of market information affecting the stock price at a certain time determines either the stock price's rise or the stock price's decline. In order to illustrate our model, by following \cite{Zhang}, we consider an idealized model in which we assume two type of information appear periodically. We choose a cosine function $\cos \omega t$ to simulate the fluctuation of the information and use the Hamiltonian
\begin{equation}
\begin{array}{l}
\hat H=\frac{1}{2\mu }\hat {\mathcal{T}}^2+\beta \, \hat {\mathcal{R}} \cos \omega t.
\end{array}
\end{equation}
where $\beta $ is a constant.
The solution of the Schr\" odinger equation 
\begin{equation}
{\rm i}\, \frac{\partial \Psi }{\partial t}=\left[ \frac{1}{2\mu }\hat {\mathcal{T}}^2+\beta \, \hat {\mathcal{R}} \cos \omega t\right] \Psi 
\end{equation}
can be obtained by using a program in Mathematica \cite{Cotfas1}.
We assume that at the opening time ($t=0$) of the stock market, the wave function describing the rate of return is the function $\Upsilon _\kappa $ (corresponding to a certain $\kappa $), that is,
\begin{equation}
\Psi (n,0)=\Upsilon _\kappa (n).
\end{equation}
The distribution of the probabilities $|\Psi (n,t)|^2$ corresponding to the possible values $n$ of the rate of return at certain moments of time are presented in Fig. 2. We can remark that, for certain moments of time ($t\!=\!28800$ seconds, in the presented example),  the previsions of our model may be ambiguous.

\section{Concluding remarks}

The quantum models facilitate the description of phenomena not fully explained by the classical models. The proposed approach takes into consideration the discrete nature of the financial variables and keeps the essential characteristics of a usual infinite-dimensional model \cite{Zhang}. \ It uses a simpler mathematical formalism: finite sums instead of integrals, operators defined on the whole Hilbert space and with finite spectrum, system of  first order differential equations instead of a second order differential equation with partial derivatives, etc.

\end{document}